\documentclass[letterpaper]{article} % DO NOT CHANGE THIS
\usepackage{aaai2027}  % DO NOT CHANGE THIS

\usepackage{times}  % DO NOT CHANGE THIS
\usepackage{helvet}  % DO NOT CHANGE THIS
\usepackage{courier}  % DO NOT CHANGE THIS
\usepackage[hyphens]{url}  % DO NOT CHANGE THIS
\usepackage{graphicx} % DO NOT CHANGE THIS
\usepackage{natbib}  % DO NOT CHANGE THIS AND DO NOT ADD ANY OPTIONS TO IT
\usepackage{booktabs}
\usepackage{tabularx}
\usepackage{array}
\usepackage{enumitem}
\usepackage{siunitx}
\usepackage[table]{xcolor}
\usepackage{caption} % DO NOT CHANGE THIS AND DO NOT ADD ANY OPTIONS TO IT
\usepackage{float}
\usepackage{amssymb}  % mathbb
\usepackage{algpseudocode}
\usepackage[boxruled,vlined,linesnumbered,algo2e]{algorithm2e}

\usepackage{longtable}
\usepackage{array}
\usepackage{booktabs}

\usepackage{tikz} % Added for tikzpicture environment
\usetikzlibrary{arrows.meta} % Enable 'Latex' arrow tip
\usepackage{amsmath}
\usepackage{colortbl}

\definecolor{GBAInk}{HTML}{303442}
\definecolor{GBABlue}{HTML}{397A8C}
\definecolor{GBAGreen}{HTML}{5D7F3E}
\definecolor{GBAGold}{HTML}{A66C16}
\definecolor{GBACoral}{HTML}{A94C46}
\definecolor{GBAPurple}{HTML}{705B8E}
\definecolor{GBALine}{HTML}{8C929E}

\newcommand{\gbasection}[2]{\arrayrulecolor{#1}\specialrule{0.7pt}{0.3ex}{0.15ex}\arrayrulecolor{GBALine}\multicolumn{3}{@{}l}{\textcolor{#1}{\bfseries #2}}\\[-0.35ex]}

\usepackage{newfloat}
\usepackage{listings}
\usepackage{booktabs} 
\usepackage{array}
\usepackage[table]{xcolor}

\usepackage{longtable}
\usepackage{array}
\usepackage{booktabs}

\newcommand{\abacus}{\textsc{ctf-abacus} }

\DeclareCaptionStyle{ruled}{labelfont=normalfont,labelsep=colon,strut=off} % DO NOT CHANGE THIS
\floatstyle{ruled}
\newfloat{listing}{tb}{lst}{}
\floatname{listing}{Listing}
\usepackage{subcaption}  % preamble

\title{How Do LLM Agents Actually Get the Flag? \\
Trace-Level Provenance for Agentic Offensive Security Evaluation}

\author{
    Kimberly Milner\textsuperscript{\rm 1}\thanks{Authors contributed equally to this research.},
Minghao Shao\textsuperscript{\rm 1,2*},
Nanda Rani\textsuperscript{\rm 3*},
Haoran Xi\textsuperscript{\rm 1},
Venkata Sai Charan Putrevu\textsuperscript{\rm 5},
Meet Udeshi\textsuperscript{\rm 1},
Sandeep K. Shukla\textsuperscript{\rm 4},
Prashanth Krishnamurthy\textsuperscript{\rm 1},
Farshad Khorrami\textsuperscript{\rm 1} \\
Muhammad Shafique\textsuperscript{\rm 2}
Ramesh Karri\textsuperscript{\rm 1}
}

\affiliations{
\textsuperscript{\rm 1}NYU Tandon School of Engineering,
\textsuperscript{\rm 2}NYU Abu Dhabi,
\textsuperscript{\rm 3}CISPA - Helmholtz Center for Information Security, Saarbrücken \\
\textsuperscript{\rm 4}International Institute of Information Technology Hyderabad, 
\textsuperscript{\rm 5}Indian Institute of Technology Tirupati\\
}

\usepackage{bibentry}
\nocopyright     
\begin{document}

    \maketitle

\begin{abstract}

\textit{Capture-the-Flag} (CTF) benchmarks are widely used to assess the offensive security capabilities of autonomous language-model agents. Evaluations rely on shallow binary judgments or aggregate scores, overlooking the agent’s trajectory to the flag. Consequently, actual exploitation is conflated with direct flag exposure, memorized recall, external lookup, guessing, and unsupported claims, potentially overstating the agent's cybersecurity capability. We introduce \textsc{ctf-abacus}, a trace-based agent auditing framework that reconstructs each run as an evidence-grounded solve profile. By decomposing agent actions into penetration-testing phases and categorical techniques, it identifies where exploitation occurs, where the flag first appears, and whether the recovered flag is supported by demonstrated behavior. Aggregating solve profiles across agents yields challenge signatures that reveal whether success was achieved via the intended exploit or via shortcut pathways. We apply \abacus to 1,435 CTF attempts by six frontier and
open-source models on 240 challenges, yielding 2,870 solve profiles under two       judge lenses.  Trace-verified exploits account for only 62-87\% of recovered flags across benchmarks, while shortcut recoveries follow substantially shallower trajectories. These findings shift CTF evaluation from counting recovered flags to verifying demonstrated exploitation and provide a basis for designing benchmarks that better isolate the offensive capabilities.

\end{abstract}

\section{Introduction}

Autonomous language-model agents can now perform complex offensive-security tasks. Recent systems plan multi-step attacks, interact with software environments, invoke external tools, and solve Capture-the-Flag (CTF) challenges that previously required experienced human participants~\cite{yang2023hackers,happe2023pwned,fang2024llm,shao2024empirical}. CTF benchmarks have consequently become a central testbed for comparing cyber agents, evaluating planning architectures, and reasoning about their security implications~\cite{caibench2025,ctfdojo2025}.

Yet most CTF evaluations reduce each run to a single outcome, whether the submitted flag matches
the ground truth, treating recovery as evidence of the targeted capability. For modern models, the same flag can arise through different pathways. A model may execute the intended exploit, read an exposed flag, recall a public solution, retrieve a write-up, guess the answer, or assert it without supporting observations.
Conventional scores collapse these into one solve, overstating capability and
obscuring what the benchmark measures, a risk amplified by contamination and memorization~\cite{dong2024contamination,ctfusion2026}. 
%A correct answer does not reveal the capability that produced it.

% Yet most CTF evaluations reduce each run to a single outcome: whether the submitted flag matches the expected answer. This treats flag recovery as evidence that the agent demonstrated the capability targeted by the challenge. For modern language models, however, the same correct flag can arise through different pathways. A model may execute the intended exploit, read an exposed flag, recall a public solution, retrieve a write-up, guess the answer, or assert it without supporting observations. Conventional scores collapse these cases into the same successful solve, potentially overstating agent capability and obscuring what the benchmark measures. Reported gaps between public benchmarks and newly released competition challenges further show that contamination and memorized solutions can materially affect evaluation~\cite{dong2024contamination,ctfusion2026}. More broadly, a correct answer does not reveal the capability that produced it.

\begin{figure}[!t]
    \centering
    \includegraphics[width=1\linewidth]{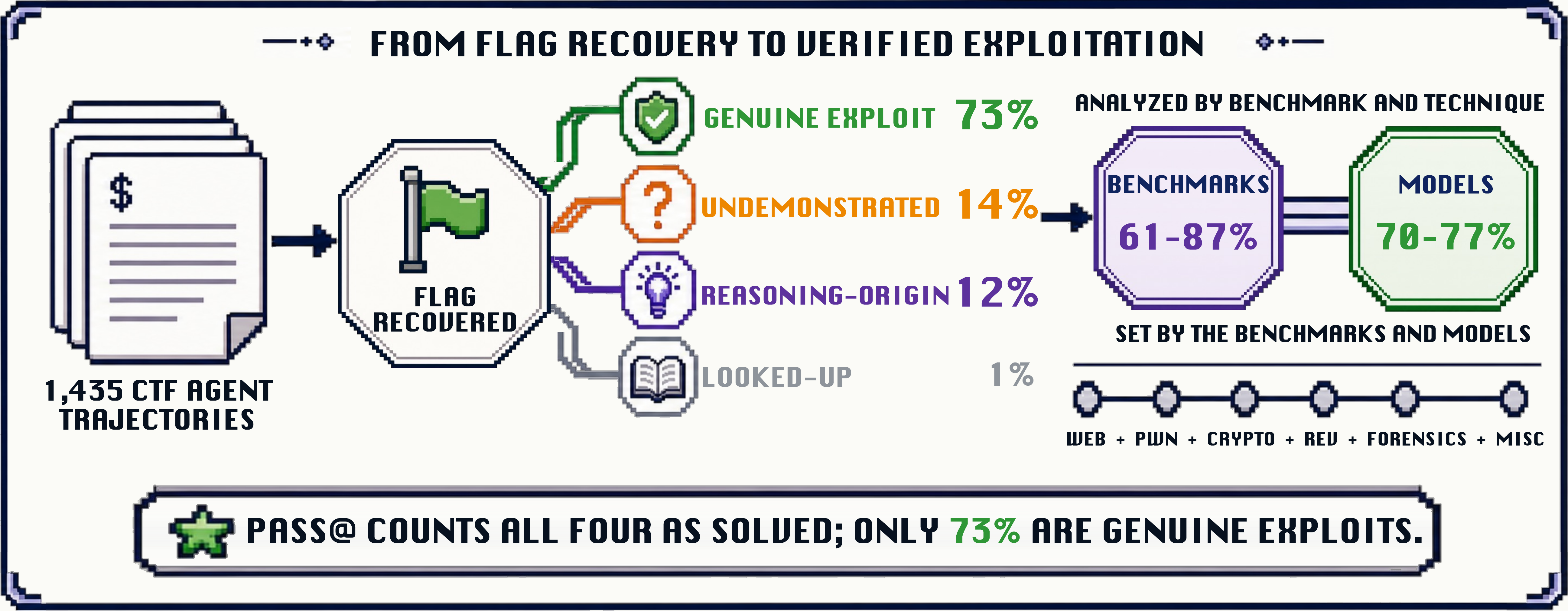}
    \caption{CTF-Abacus reconstructs each run from its execution trace, separating demonstrated exploitation from shortcuts that flag matching counts as solved.}
    \label{fig:teaser}
\end{figure}

% Evaluating agents on unseen competitions and improving benchmark construction can reduce some of these risks~\cite{ctfusion2026,ctfdojo2025,happe2023pwned}. They nevertheless leave a more fundamental question unanswered: \emph{when an agent submits a correct flag, what evidence shows that it actually performed the intended exploit?} Existing benchmarks verify the answer, but not the execution that produced it. We argue that this evidence lies in the execution trace. We introduce \textsc{ctf-abacus}, an automated framework that reconstructs each run as an evidence-grounded \emph{solve profile}. It decomposes agent actions into penetration-testing phases and category-specific security techniques, identifies where exploitation occurs and where the flag first appears, and determines whether the recovered flag is supported by demonstrated behavior. This distinguishes genuine exploitation from exposed flags, memorization, lookup, guessing, unsupported claims, and other shortcuts. Aggregating profiles across agents and models yields \emph{challenge signatures} that reveal whether success depends on the capability a challenge is intended to test.

Evaluating agents on unseen challenges and improving benchmark construction reduce these risks~\cite{ctfusion2026,ctfdojo2025,happe2023pwned}, but leave a fundamental question: \emph{when an agent submits a correct flag, what evidence shows that it actually performed
the intended exploit?} Existing benchmarks verify the answer, but not the execution that produced
it. This evidence lies in the execution trace. We introduce \textsc{ctf-abacus}, an automated framework that reconstructs each run as an evidence-grounded
\emph{solve profile}: it decomposes agent actions into penetration-testing phases and
category-specific security techniques, locates where exploitation occurs and where the flag first appears, and determines whether the recovered flag is supported by demonstrated behavior. This
separates exploitation from shortcuts like memorization and flag lookups. Aggregating profiles across models and categories yields \emph{challenge signatures}
that reveal whether success depends on the intended capability.

% Validating such judgments ultimately requires inspecting the trace itself, which has not scaled
% to thousands of multi-step runs. By making that inspection tractable our observability effort
% turns the distinction between genuine exploitation and mere flag recovery from an assertion into
% a validated measurement, and points toward a general means of making agent behavior auditable.

We use \textsc{ctf-abacus} not only to audit individual attempts but to study autonomous CTF agents and the
benchmarks that assess them: how agents progress through penetration-testing phases, which
techniques they employ, and where successful trajectories diverge from demonstrated exploitation.
The analysis surfaces recurring structures separating genuine exploits from shortcut
recoveries showing conventionally equivalent challenges can demand different offensive capabilities. \textsc{ctf-abacus} thus turns execution traces into an empirical instrument for
studying agent behavior and benchmark validity. Our contributions are as follows:

\begin{enumerate}[leftmargin=*, label=\textbullet]
\item We introduce an automated trace-based framework with a observability instrument transforming traces into evidence-grounded solve profiles, attributing recovered flags to demonstrated exploitation or alternative pathways.

\item We develop a systematic method for auditing challenge validity by aggregating solve profiles into challenge signatures, revealing how strongly success depends on the capability each challenge is intended to measure.

\item We conduct a comprehensive large-scale study of 1,435 attempts, six frontier language models, and 240 challenges from four CTF benchmarks, yielding over 2,870 solve profiles under two independent judges. Trace-verified exploitation validates only 62-87\% of recovered flags, showing that flag-based scores can obscure substantial differences in agent behavior and challenge validity.
\end{enumerate}

\section{Related Work}

\subsection{CTFs as the Unit of Evaluation}

InterCode-CTF introduced the first interactive CTF environment for language agents, comprising 100 picoCTF tasks in a Bash shell. GPT-4 solved 40 tasks but struggled with challenges requiring multi-step investigation \cite{yang2023hackers}. Work on competition-grade challenges followed, comparing
fully automated runs against human-assisted ones \cite{shao2024empirical}.
Related systems moved past puzzle challenges, exploiting live web applications
without human guidance \cite{fang2024llm} and producing both attack plans and the
shell commands that carry them out \cite{happe2023pwned}. Recent environments are
much larger. CAIBench spans over 10,000 instances across Jeopardy CTFs,
attack-and-defense rounds, cyber ranges, knowledge tests, and privacy tasks, and
reports that security knowledge held by a model does not translate into attack or
defense ability \cite{caibench2025}. CTF-Dojo packages 658 containerized
challenges generated automatically from public artifacts, then trains agents on
486 execution-verified trajectories for gains of up to 11.6\% absolute
\cite{ctfdojo2025}. That last step is where our concern becomes concrete. A
trajectory is verified by the flag it produced, so the path taken to that flag is
never examined before it becomes a training signal.

\subsection{Validity Critiques and Contamination Evidence}

How these agents are scored has already drawn criticism. A systematic review of
LLM-driven offensive-security prototypes reports that most publish little beyond
success counts, that testbeds and experimental designs differ too widely to
support comparison, and that the field needs baselines, richer metrics, and
qualitative analysis of agent behavior \cite{happe2025benchmarking}.
Contamination research raises the same doubt from the data side: memorized
benchmark content inflates measured performance and is readily mistaken for
reasoning, yet the detectors built for it read benchmark-level output statistics
and return no verdict on an individual solve \cite{dong2024contamination}.
CTFusion quantifies the effect in CTFs, reporting agent performance on the reused
NYU CTF Bench (14.4\%) well above performance on live competition challenges
(6.3\%), and showing that adding web search to an agent lifted its score from
12.59\% to 24.07\% while several of its submitted flags came from public solutions \cite{ctfusion2026}. These results characterize populations of attempts but do not reveal which solves were genuinely earned, a distinction execution traces can establish.

\subsection{Behavioral Analysis of Execution Traces}

Closest to our approach is work that inspects agent behavior rather than scores.
CyBiasBench records 630 sessions from five agents against three targets under
four prompt conditions, and shows that each agent concentrates on a narrow subset
of ten attack families no matter how it is prompted \cite{cybiasbench2026}. That
analysis concerns preference, meaning which attacks an agent reaches for. Ours
concerns warrant, meaning whether the actions recorded in a trace support the
flag that was submitted. The difference matters because a biased agent can still
solve a challenge honestly, and an unbiased one can still submit a recalled flag.
To our knowledge, no prior work provides per-attempt evidence that the intended exploitation occurred or aggregates such evidence across models to measure what a benchmark actually tests.
% To our knowledge, no prior work derives per-attempt evidence of whether the
% intended exploitation occurred, or aggregates that evidence across models into a
% challenge-level measure of what a benchmark actually tests.

\section{Methodology}
\label{sec:method}

% Flag recovery alone does not establish that an agent demonstrated the capability a CTF challenge
% targets. A correct flag may follow from exploiting the intended weakness, but it can equally arise
% from recollection, external lookup, a lucky guess, or an unsupported assertion. We therefore
% evaluate each attempt by its \emph{flag provenance}: where the flag came from, and whether it was
% earned. From an agent's execution trace we reconstruct a \emph{solve profile}; assign the flag's
% provenance under two independent model lenses; audit that reconstruction against the underlying
% evidence; require human domain experts to review the cases the trace cannot resolve by examining pure tool outputs; and aggregate profiles across
% models into a \emph{challenge signature} (Figure~\ref{fig:framework}). The rest of this section
% develops each stage.

% A recovered flag need not prove capability; it may be recalled, looked up, guessed, or merely
% asserted~\cite{ctfusion2026}. We judge each attempt by its flag provenance $\pi$: where it came
% from and whether it was earned. From the trace we reconstruct a \emph{solve profile}, assign
% $\pi$ under two model lenses, audit it, route unresolved cases to human experts, and aggregate
% into a \emph{challenge signature} (Figure~\ref{fig:framework}).

\begin{figure*}[t]
  \centering
  \includegraphics[width=0.97\linewidth]{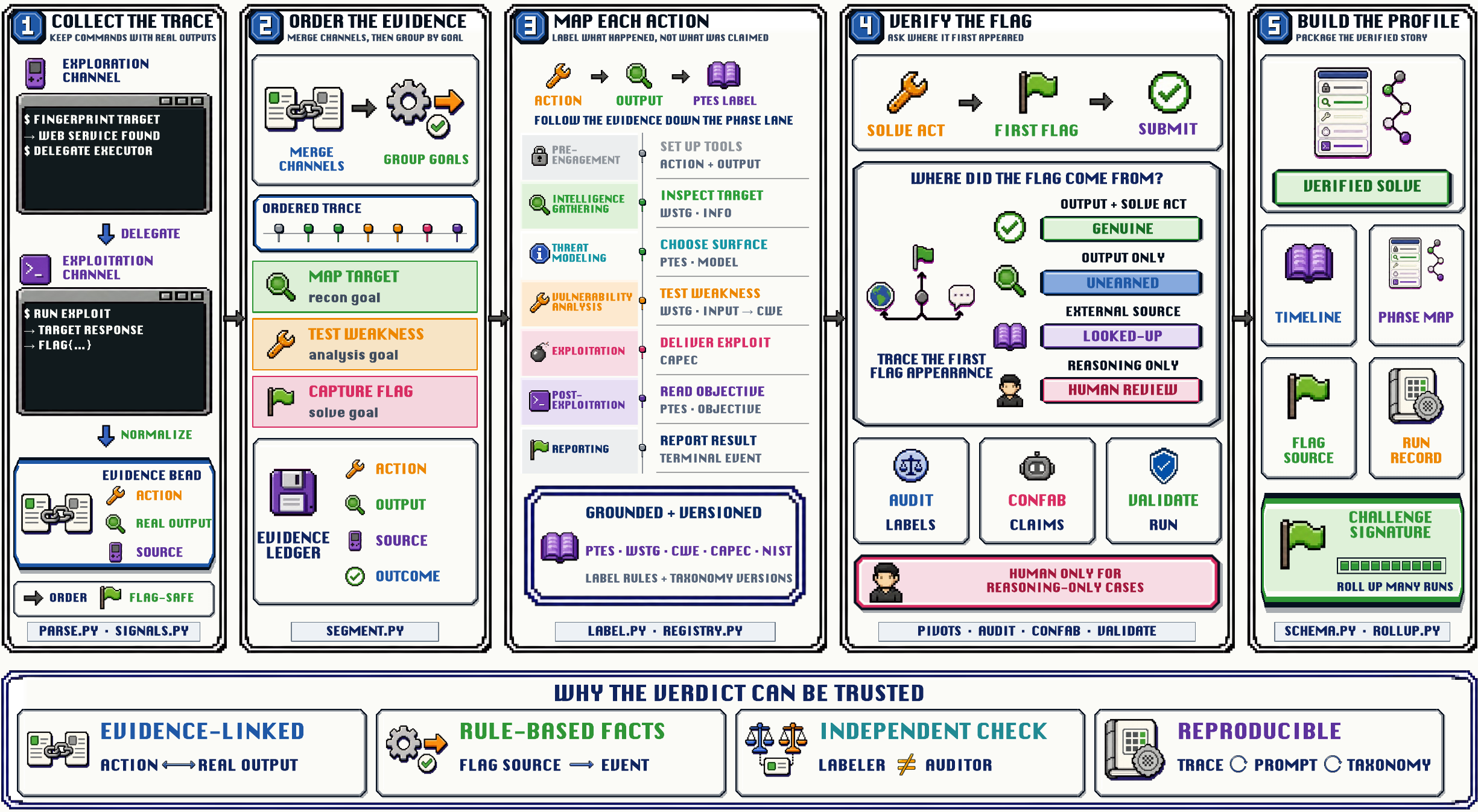}
  \caption{\footnotesize\textbf{From execution trace to verified solve profile.} Multi-agent trajectories are linearized into steps $(a_i,o_i)$, labeled with penetration-testing phases $\phi_i$ and techniques $\tau_i$, assigned flag provenance $\pi$, evidence-audited, and aggregated into challenge signatures. Human review is restricted to reasoning-originated flags classified as \textit{recalled, guessed, leaked, or derived}.}
  \label{fig:framework}
\end{figure*}

A recovered flag does not necessarily prove the capability; it may result from exploitation, recall, external lookup, guessing, or an unsupported claim~\cite{ctfusion2026}. We judge each attempt by its flag provenance $\pi$: where it came from and whether it was earned. From the trace we reconstruct a \emph{solve profile}, assign provenance under two  model lenses, audit it against evidence,  route unresolved cases to human experts, and aggregate profiles into a \emph{challenge signature}
(Figure~\ref{fig:framework}). We discuss each stage below.

\begin{table}[t]
  \centering
  \fontsize{7}{7.8}\selectfont
  \color{GBAInk}

  \setlength{\tabcolsep}{2.2pt}
  \renewcommand{\arraystretch}{0.95}
  \setlength{\extrarowheight}{0pt}

  \setlength{\aboverulesep}{0.15ex}
  \setlength{\belowrulesep}{0.15ex}
  \setlength{\abovecaptionskip}{0pt}
  \setlength{\belowcaptionskip}{1.5pt}

  \caption{Solve-profile notation and flag-provenance rules.}
  \label{tab:notation}
  % \vspace{-0.1mm}

  \begin{tabularx}{\columnwidth}{
    @{}
    >{\raggedright\arraybackslash}p{0.27\columnwidth}
    >{\centering\arraybackslash}p{0.18\columnwidth}
    >{\raggedright\arraybackslash}X
    @{}
  }
    \arrayrulecolor{GBAInk}
    \toprule
    \rowcolor{GBAInk!6}
    \textbf{Item} & \textbf{Symbol} & \textbf{Meaning} \\
    \gbasection{GBABlue}{Trace notation}
    Execution trace & $T=(e_1,\ldots,e_n)$ & Ordered sequence of $n$ execution steps. \\
    \rowcolor{GBAInk!3}
    Step & $e_i=(a_i,o_i)$ & Agent action $a_i$ and corresponding observation $o_i$. \\
    Step labels & $z_i=(\phi_i,\tau_i)$ & PTES phase $\phi_i$ and standards-grounded technique $\tau_i$. \\

    \gbasection{GBAGreen}{PTES depth}
    \rowcolor{GBAInk!3}
    Phase order & $1\rightarrow5$ & Pre-engagement $\rightarrow$ Intelligence Gathering $\rightarrow$ Vulnerability Analysis $\rightarrow$ Exploitation $\rightarrow$ Post-Exploitation. \\

    \gbasection{GBAGold}{Pivotal events}
    Exploitation act & $\xi^\dagger$ & First genuine target exploit at Vulnerability Analysis depth or deeper. \\
    \rowcolor{GBAInk!3}
    First appearance & $f,\sigma$ & First flag appearance $f$ and its source $\sigma$. \\
    Submission & $s$ & First submitted flag, correct or not. \\
    \rowcolor{GBAInk!3}
    Solve profile & $P(T)$ & $P(T)=\bigl((e_i,z_i)_{i=1}^{n},\xi,f,\sigma,s\bigr)$ \\

    \gbasection{GBACoral}{Flag provenance $\pi$}
    genuine-solve & $f$ after $\xi$ & Flag observed from the target after demonstrated exploitation. \\
    \rowcolor{GBAInk!3}
    undemonstrated exploit & target, no $\xi$ & Target flag appears without a demonstrated exploit. \\
    looked-up & external & Flag obtained from an external source. \\
    \rowcolor{GBAInk!3}
    agent-reasoning & reasoning & Flag appears only in the agent's reasoning. \\
    unsolved & no $f$ & No flag appears in the run. \\

    \gbasection{GBAPurple}{Human refinement}
    \rowcolor{GBAInk!3}
    recalled & training & Memorized from training data or prior exposure. \\
    leaked & run context & Surfaced from the run context rather than training. \\
    \rowcolor{GBAInk!3}
    guessed & guess & Produced through a lucky guess. \\
    derived & reasoning & Computed in reasoning without a supporting observation. \\
    \rowcolor{GBAInk!3}
    unclear & unresolved & Evidence is insufficient to determine the source. \\
    \arrayrulecolor{GBAInk}
    \bottomrule
  \end{tabularx}
  \vspace{-0.9mm}
  \parbox{\columnwidth}{\scriptsize\raggedright $^\dagger\xi$ is adjudicated by one reasoning-aware pass; $f$, $\sigma$, and $s$ are read directly from the trace. $\sigma\in\{\text{environment},\text{external},\text{reasoning}\}$.\par}
  \vspace{-2.3mm}
\end{table}

\paragraph{Solve-profile processing.} The \textsc{D-CIPHER}~\cite{udeshi2025dcipher} run
interleaves a planner with executor sub-agents; we flatten it into a time-ordered sequence $T=(e_1,\ldots,e_n)$, splicing each executor's steps in at the planner delegation that spawned them. Each step $e_i=(a_i,o_i)$ is a tool call $a_i$ with an observation
window $o_i$; routing-only control operations, acting on no target, are dropped. Reasoning messages are not steps but are for provenance, as a flag may first surface there.

%The window keeps roughly the first and last $8{,}000$ characters and is flag-aware: if the true flag would fall in the truncated middle, the window shifts to retain it.
Each step then carries a label $z_i=(\phi_i,\tau_i)$: a PTES~\cite{ptes} phase $\phi_i$ and a
technique $\tau_i$ from a category-specific standard, OWASP WSTG and the Top-10 for
web~\cite{owasp-wstg,owasp-top10}; CWE, CAPEC, and MITRE ATT\&CK for binary exploitation and
reverse engineering~\cite{cwe,capec,mitre-attack}; NIST for cryptography and
forensics~\cite{nist800115,nist80086}. The registry is open-set: unmatched actions generate candidate techniques, which are added after human verification and alignment with an existing standard.

% \paragraph{Solve-profile processing.} The \textsc{D-CIPHER}~\cite{udeshi2025dcipher} run
% interleaves a planner with executor sub-agents; we flatten it into a time-ordered sequence
% $T=(e_1,\ldots,e_n)$, splicing each executor's steps in at the planner delegation that
% spawned them. Each step $e_i=(a_i,o_i)$ is a tool call $a_i$ with a bounded observation
% window $o_i$; routing-only control operations, acting on no target, are dropped.
% Reasoning messages are not steps but are kept for provenance, as a flag may first surface there.

% %The window keeps roughly the first and last $8{,}000$ characters and is flag-aware: if the true flag would fall in the truncated middle, the window shifts to retain it.

% \paragraph{Solve-profile labeling.} Each step $e_i$ carries a label $z_i=(\phi_i,\tau_i)$:
% a PTES~\cite{ptes} phase $\phi_i$ and a technique $\tau_i$ from a category-specific standard,
% OWASP WSTG and the Top-10 for web~\cite{owasp-wstg,owasp-top10}; CWE, CAPEC, and MITRE
% ATT\&CK for binary exploitation and reverse engineering~\cite{cwe,capec,mitre-attack}; NIST
% for cryptography and forensics~\cite{nist800115,nist80086}. The registry is open-set: an
% action matching no existing technique prompts a suggested new $\tau_i$, promoted into the
% taxonomy on human verification and mapping to an existing cybersecurity standard. 

\paragraph{Flag provenance and validation.} Whether a flag appeared is trivial; \emph{how} it
arrived is decisive. We read its first appearance $f$ (source $\sigma$) and correct submission
$s$ deterministically; only the \emph{exploitation act} $\xi$, the first step that genuinely
exploits the target, is inferred. Lacking reference solutions, $\xi$ is anchored in
demonstrated depth, not a known path, and a single reasoning-aware pass discards test-flag,
scaffold, and recalled reads. Provenance $\pi$ then follows: a target-observed flag is a
\emph{genuine-solve} if $\xi$ preceded it, else an \emph{undemonstrated exploit}; a
reasoning-only flag is \emph{agent-reasoning}. We report provenance, not flag presence, as
a solve. Because $\xi$ is inferred, we validate rather than assert it: two independent judges,
Claude and GPT, label from the action and observation alone, so running
both keeps either from favoring a solver of its own family. A blind, family-matched audit then scores each reconstruction against the trace (0 to 100) on how well its phase, technique, and  flag-event tags fit evidence; this fidelity averages \num{98.3}/100. To validate these labels, security researchers re-judged a 10\% sample ($N=92$) with $\pi$ verdict and all pivot events hidden, agreeing with the pipeline on 93.5\% of genuine solves and 86.7\% of undemonstrated exploits. Cases that resist deterministic resolution are  human-adjudicated: all \num{127} \emph{reasoning-origin}
flags (Table~\ref{tab:notation}: \textit{Human Refinement}) resolve to
\emph{derived} (\num{115}), \emph{recalled} (\num{9}), \emph{leaked} (\num{1}),
\emph{guessed} (\num{1}), or \emph{unclear} (\num{1}).

\section{Experiment Setup}

% figure: figures/results-fig-geometry/  (make.py -> data.csv + results-fig-geometry.pdf)

\begin{figure*}[!t]
  \centering
  \includegraphics[width=\linewidth]{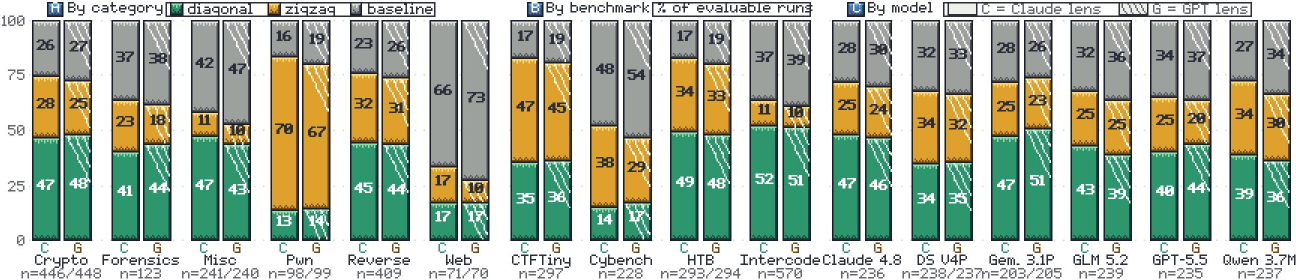}
  \caption{\footnotesize\textbf{Trajectory geometry across category, benchmark, and model, under both judge lenses.}
  Diagonal, zigzag, and baseline composition of all evaluable runs, cut three ways. Each group shows the
  Claude lens (C, upper bar) over the GPT lens (G, lower bar). Composition shifts sharply across category
  and benchmark but barely across model; the two lenses track closely.}
  \label{fig:geometry}
\end{figure*}

\paragraph{Benchmarks Evaluated.} We evaluate the agents on four widely used offensive-security benchmarks: Hack The Box (HTB)~\cite{abramovich2025enigma}, InterCode-CTF~\cite{yang2023intercode}, CTFTiny~\cite{shao2025ctftiny}, and CyBench~\cite{zhang2024cybench} with 240 challenges in total. These benchmarks span common CTF categories such as web exploitation, binary exploitation, reverse engineering, cryptography, forensics, and miscellaneous security tasks, while differing in challenge construction, execution environment, and interaction complexity. 
Evaluating across known benchmarks tests whether observed solve behaviors and benchmark-level patterns persist across diverse settings, %rather than reflect a single benchmark, 
strengthening the breadth and generalizability of our analysis.
% Evaluating across these established benchmarks allows us to examine whether the observed solve behaviors and benchmark-level patterns persist across diverse settings rather than reflecting the characteristics of a single benchmark, thereby strengthening the comprehensiveness of our evaluation and the generalizability of the resulting analysis.

\paragraph{Models and Environment Setup.} We evaluate six models, Claude Opus 4.8, GPT 5.5, Gemini 3.1 Pro, GLM 5.2, DeepSeek V4 Pro, and Qwen 3.7 Max, selected as among the strongest available at evaluation time, so our findings reflect capable agents rather than a single lineage. Two judges from different families, Claude Opus 4.8 and GPT 5.6 Sol, evaluate traces, reducing model-specific bias. Each agent runs under the harness's default configuration; following prior offensive-security evaluations, we cap model-API cost at \$3 per attempt, beyond which an attempt counts as failed.

\section{Evaluation Results}
\label{sec:evaluation}

% We evaluate \textsc{CTF-ABACUS} on 2{,}870 attempts generated by two solver harnesses, \textsc{D-CIPHER}~\cite{udeshi2025dcipher} and \textsc{ENIGMA}~\cite{abramovich2024enigma}. 
We evaluate \textsc{CTF-ABACUS} over 1400 \textsc{D-CIPHER} attempts from six models across 240 challenges and four benchmarks. After excluding 47 non-evaluable runs 1,056 flag recoveries are profiled with each evaluable attempt assessed independently under the Claude and GPT framework lenses. Unless stated, results presented use the Claude lens. \emph{Recovery} means a valid flag appears, whereas \emph{execution-backed} requires an observed attack. A recovery without observed execution may be \emph{derived} when its reasoning is human-verified; the remainder are unsupported. Verified scores use the execution-backed criterion. We use these traces to assess whether recovered flags are execution-backed, how verification affects scores, and which behaviors distinguish recovery pathways.

\paragraph{Anatomy of a Solve Profile}

CTF trajectories are visualized through their measure of \emph{monotonicity}, the fraction of PTES $\phi$ transitions that advance the attack towards advanced PTES stages. The value is used to construct geometries with \emph{diagonal} (a steady climb, \textit{mono} $\geq 0.8$), \emph{zigzag} (net progress with backtracking, \textit{mono} $\geq 0.5$), and \emph{baseline} (no net
progress, i.e., a grind). 
% Shape is governed by the challenge, not the model. It is nearly constant
% across the six models but swings across categories and benchmarks, from 70\% zigzag in binary
% exploitation to 48\% baseline in Cybench. Fig.~\ref{fig:geometry} quantifies this. 
We find geometry ties weakly to the model (Cram\'er's $V$ of 0.08/0.10) and about $3\times$ as strongly to category (0.26/0.27) and benchmark (0.29/0.28), for both lenses (Claude, GPT).
%The two cross model judge lenses also agree on the geometries of 83\% of runs (Cohen's $\kappa=0.75$), with disagreements concentrated at the climbing-flat boundary rather than between climbing geometries (Table~\ref{tab:geometry}). 
Hence we find genuine-exploit $\pi$  for a solver requires the agent to achieve more advanced PTES $\phi$ (even with zig-zagging regressive steps), than in simpler challenges where flags can just be read from the environment or asserted flatten trajectories.

\paragraph{Cyber Taxonomies}

Aligning every step $e$ to  finer grained cybersecurity standardization and techniques $\tau$ provides insight into agent activity distribution across the solve. (Tab.~\ref{tab:PTES_effort_r}). The trend emerges of  solve profiles being heavily front-loaded in PTES $\phi$=\textit{Intelligence Gathering}. The two PTES  $\phi$ that bear on genuine capability, Vulnerablity Analysis and Exploitation, strongly constrast in $\tau$ profiling. Vulnerability Analysis carries a distinct technique signature as Exploitation \emph{converges} (across every benchmark and model) reduces to authoring and running a solve script (excluding Web, which alone stays domain-specific. Where Vulnerability Analysis fingerprints the domain, Exploitation erases it wherever the flag can be reached by running code.

\begin{table}[htbp]
  \centering
  \scriptsize
  \setlength{\abovecaptionskip}{3pt}   % caption sits close to top
  \setlength{\belowcaptionskip}{1pt}   % <- kills the title-to-table gap
  \setlength{\tabcolsep}{3.8pt}
  \renewcommand{\arraystretch}{0.8}
  \caption{Effort and technique diversity by PTES $\phi$}
  \label{tab:PTES_effort_r}
  \begin{tabularx}{\columnwidth}{@{}%
    >{\raggedright\arraybackslash}p{0.28\columnwidth}%
    >{\centering\arraybackslash}p{0.20\columnwidth}%
    >{\raggedright\arraybackslash}X@{}}
    \toprule
    \textbf{Phase} & \textbf{Steps / \#\(\tau\) (\textit{Eff})} & \textbf{Dominant technique (share)} \\
    \midrule
    Reconnaissance         & 52\% / 71\,(12)  & Probe live service (27\%) \\
    Vulnerability analysis & 19\% / 126\,(42) & Category-specific ($\le$18\%) \\
    Exploitation           & 15\% / 66\,(13)  & Author solve script (40\%) \\
    Post-exploitation      & $<$2\% / 12\,(2) & Read captured flag (80\%) \\
    \bottomrule
  \end{tabularx}
  \vspace{-1mm}
  {\footnotesize\raggedright \emph{Steps}: fraction of agent steps $e$ in $\phi_i$; \textit{Eff}: effective \#$\tau$ in $\phi_i$.\par}
\end{table}

% \paragraph{Behavioral depth of genuine solves.}

\paragraph{Behavioral depth of genuine solves.}
Genuine solves and shortcut recovery profiles  exhibit starkly different geometries: genuine solves climb the PTES $\phi$ to max depth, monotonically or with some zig-zagging. Shortcut recoveries grind, producing a flag without ever exhibiting exploitation traces in tool outputs (Table~\ref{tab:genuine_depth}).

\begin{table}[t]
  \centering
  \small
  \setlength{\abovecaptionskip}{2pt}
  \setlength{\belowcaptionskip}{0pt}
  \setlength{\tabcolsep}{4pt}
  \renewcommand{\arraystretch}{0.8}
  \setlength{\aboverulesep}{1pt}
  \setlength{\belowrulesep}{1pt}
  \setlength{\heavyrulewidth}{.06em}
  \setlength{\lightrulewidth}{.04em}
  \caption{\footnotesize{genuine \& $\neg\phi$=Exploit Flag recoveries.}}
  \label{tab:genuine_depth}
  \begin{tabular}{@{}lcccc@{}}
    \toprule
     & \textbf{\# $\tau$} & \textbf{PTES depth} & \textbf{monotonicity $m$} & \textbf{reached Exp.} \\
    \midrule
    Genuine  & 5.2 & 4.3 & 0.80 & 89\% \\
    Shortcut & 4.3 & 2.8 & 0.50 & 42\% \\
    \bottomrule
  \end{tabular}
\end{table}

% \begin{table}[htbp]
%   \centering
%   \small
%   \setlength{\abovecaptionskip}{3pt}
%   \setlength{\belowcaptionskip}{1pt}
%   \setlength{\tabcolsep}{4pt}
%   \renewcommand{\arraystretch}{1.15}
%   \caption{\footnotesize{genuine \& $\neg$ $\phi$=Exploit Flag recoveries.}}
%   \label{tab:genuine_depth}
%   \begin{tabular}{@{}lcccc@{}}
%     \toprule
%      & \textbf{\# $\tau$} & \textbf{PTES depth} & \textbf{monotonicity $m$} & \textbf{reached Exp.} \\
%     \midrule 
%     Genuine  & 5.2 & 4.3 & 0.80 & 89\% \\
%     Shortcut & 4.3 & 2.8 & 0.50 & 42\% \\
%     \bottomrule
%   \end{tabular}
% \end{table}

\paragraph{Reaching $\phi$=Exploit in CTF Benchmarks}

\begin{table}[htbp]
  \centering
  \small
  \setlength{\abovecaptionskip}{3pt}
  \setlength{\belowcaptionskip}{1pt}
  \setlength{\tabcolsep}{5pt}
  \renewcommand{\arraystretch}{0.8}
  \caption{\textbf{Benchmark quality by dimension}, median over 6 models. \footnotesize{$\tau$ techniques;
  $m$ monotonicity (0--1); $d$ PTES depth (0--5); \emph{gen.}\ genuine share of recoveries;
  \emph{r.Exp}/\emph{rec.}/\emph{r.Post}\ \% of runs reaching Exploitation / recovering /
  reaching Post-Exploitation}}
  % \caption{\textbf{Benchmark quality by dimension}, median over the six models.
  % \footnotesize{$\tau$: distinct techniques; $m$: monotonicity (0--1); $d$: PTES depth
  % (0--5). \emph{gen.}: genuine share of recovered flags; \emph{r.Exp}, \emph{rec.},
  % \emph{r.Post}: \% of runs reaching Exploitation, recovering a flag, and reaching
  % Post-Exploitation.}}
  \label{tab:benchmark_quality}
  \begin{tabular}{@{}lccccccc@{}}
    \toprule
     & $\tau$ & $m$ & $d$ & gen. & r.Exp & rec. & r.Post \\
    \midrule
    CTFTiny   & 6.5          & 0.75          & \textbf{4.21} & \textbf{86} & 88          & 86          & \textbf{40} \\
    Cybench   & \textbf{9.0} & 0.50          & 3.58          & 62          & 70          & 36          & 16 \\
    HTB       & 5.5          & \textbf{0.79} & 4.13          & 81          & \textbf{91} & \textbf{87} & 22 \\
    Intercode & 2.5          & 0.65          & 3.60          & 64          & 64          & 79          & 23 \\
    \bottomrule
  \end{tabular}
\end{table}

% Ranking benchmarks by flag recovery alone misleads (Table~\ref{tab:benchmark_quality}).
% The two security suites, \textsc{HTB} and \textsc{CTFTiny}, elicit the most genuine and deepest solving,
% leading on genuine-exploit share, PTES depth, monotonicity, and reach into Exploitation
% and Post-Exploitation. \textsc{Cybench} is where agents most visibly struggle, lowest on recovery,
% genuine share, depth, and monotonicity, yet highest on technique count $\tau$: extensive
% but unsuccessful exploration, not skill. \textsc{Intercode} inverts it, high recovery yet the
% fewest techniques and the shallowest reach into Exploitation, flags recovered without
% demonstrated exploitation. Neither $\tau$ nor recovery tracks capability; the dimensions
% that bear on genuine exploitation order the benchmarks the opposite way from both.

% \begin{figure}[t]
%   \centering
%   \includegraphics[width=0.85\linewidth]{images/results-fig-tau-regime.png}
%   \caption{\footnotesize\textbf{Technique breadth is not capability.} Ordered by
%   median $\tau$, genuine share peaks at moderate breadth, not at the extremes.}
%   \label{fig:tau_regime}
% \end{figure}

\begin{figure}[t]
  \centering
  \includegraphics[width=\linewidth]{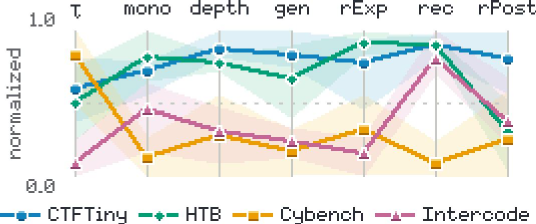}
  \caption{\textbf{Benchmark decomposition}
  {\footnotesize Axes: $\tau$ techniques; (mono) monotonocity rate, (\textit{depth}) (PTES); (\textit{gen}) genuine solve share; (\textit{rExp}) reach Exploitation; (\textit{rec}) recovery, (\textit{rPost}) reach Post-Exploitation. 
  Color bands shows per-model min-max; no clipping.}}
  \label{fig:quality}
\end{figure}

Flag recovery ranks benchmarks poorly (Table~\ref{tab:benchmark_quality},
Fig~\ref{fig:quality}). \textsc{HTB} and \textsc{CTFTiny} lead on genuine share, depth,
monotonicity, and reach; \textsc{Cybench} is lowest on all four yet highest on $\tau$
(effort, not skill); \textsc{Intercode} recovers often but shallowly. Neither $\tau$ nor
recovery tracks capability, so decomposing solves gives the yardstick a leaderboard cannot:
judge a suite by the rigor of its exploitation, not the flags it yields.
\paragraph{Technique breadth is not capability.}
Ordered by median $\tau$, capability peaks in the middle, not at the extremes
%(Fig~\ref{fig:tau_regime}). 
\textsc{Cybench}'s widest repertoire ($\tau{=}9$) solves least
(36\% recovery, 62\% genuine), thrashing, not skill; \textsc{Intercode} ($\tau{=}2.5$)
recovers often but shallowly. Capability lives at moderate breadth (\textsc{CTFTiny},
\textsc{HTB}), enough technique to carry the exploit, not so much that the agent is lost.

% \begin{figure}[t]
%   \centering
%   \includegraphics[width=\linewidth]{images/results-fig-benchmark-quality.png}
%   \caption{Benchmark quality across seven higher-is-better dimensions (Claude lens):
%   line $=$ benchmark median over six models, band $=$ per-model min--max, each axis spans
%   the min-to-max of any single model.}
%   \label{fig:quality}
% \end{figure}

\begin{figure}[t]
  \centering
  \includegraphics[width=\linewidth]{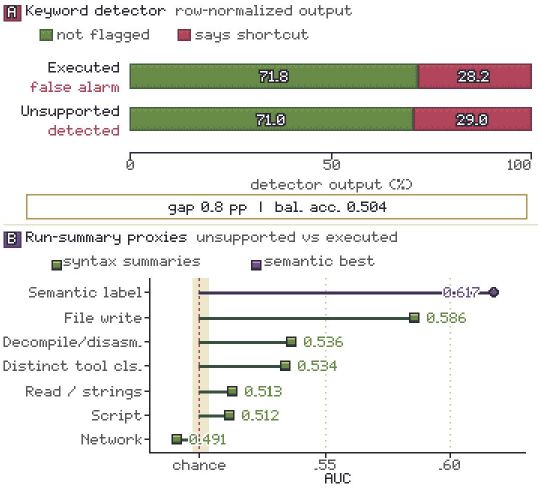}
  \caption{Command- and run-summary features poorly distinguish execution-backed from unsupported flag recoveries.}
  \label{fig:is1}
\end{figure}

\section{Discussion and Insights}
\label{sec:results}

\paragraph{No Cheap Proxy} 

% A natural question is whether trace reconstruction can be replaced by a simpler detector. We test the most direct version: locate the first appearance of the flag by exact string matching, then infer its provenance from keywords in the command that produced it. As shown in Figure~\ref{fig:is1}, this detector achieves an F1 score of only 0.28. More revealingly, it labels 28.2\% of execution-backed recoveries and 29.0\% of unsupported recoveries as shortcuts. In other words, its predictions are almost equally frequent for the two groups it is intended to distinguish. This ambiguity is inherent to the action rather than an artifact of a particular rule. A command such as \texttt{strings ./binary} is ordinary tradecraft in reverse engineering, but the same command may also reveal a flag that is directly exposed in an artifact. The command becomes meaningful only when interpreted together with what the agent had already established and what the environment returned.

We ask whether execution-backed and unsupported flag recoveries can be distinguished without reconstructing the full trace. The direct detector in Figure~\ref{fig:is1} locates the flag's first appearance and classifies the producing command using keywords. It achieves an F1 of 0.28 and flags 28.2\% of execution-backed recoveries and 29.0\% of unsupported recoveries, offering almost no separation. The same action, such as \texttt{strings ./binary}, may reflect legitimate analysis or reveal an exposed flag; its meaning depends on the preceding evidence and environment response.

Run-level summaries lack context: tool-composition yields 0.491-0.586 AUCs, while the best semantic feature reaches only 0.617. Provenance thus demands ordered relationship between actions and observations, ignored by isolated commands and scalar summaries. Motivating \textsc{CTF-ABACUS}'s sequence-level reconstruction.

% Summarizing the entire run does not recover this missing context. An AUC of 0.5 represents random discrimination, yet all tool-composition features remain between 0.491 and 0.586. Even the seemingly intuitive share of reading and \texttt{strings} steps reaches only 0.513, because genuine reverse-engineering and cryptographic work naturally contain many such actions. Replacing syntactic tool categories with semantic technique labels raises the best single-feature AUC only to 0.617. The distinguishing evidence is present in the run but lost when compressed into isolated commands or scalar summaries. This motivates \textsc{CTF-ABACUS}'s sequence-level reconstruction.

\begin{figure}[t]
  \centering
  \includegraphics[width=\linewidth]{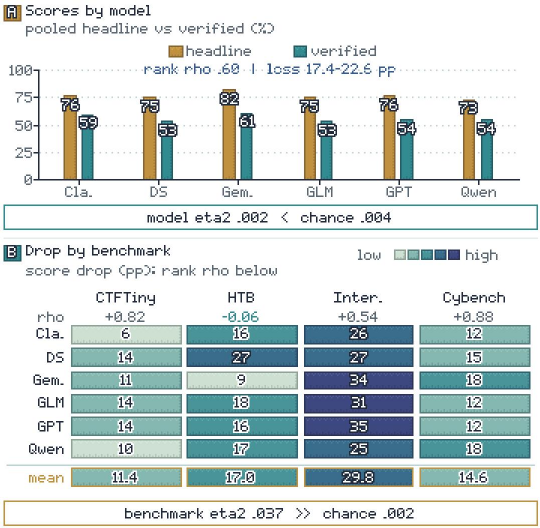}
  \caption{Verification shifts model scores similarly, while the adjustment varies more across benchmarks, making rank changes primarily benchmark-dependent.}
  \label{fig:is2}
\end{figure}

\paragraph{Ranking Instability}

% A headline leaderboard counts every recovered flag as a solve, whereas the verified score retains only flags supported by executed attack. Applying this distinction to the same runs (Figure~\ref{fig:is2}) lowers every model by 17.4--22.6\% and reduces the pooled rank correlation to 0.60. Agreement also varies by benchmark, from 0.83 on CTFTiny to -0.09 on HTB. Thus, changing the evidence required for a solve can change the apparent ordering without being driven by any single model.

Headline scores credit every recovered flag, whereas verified scores require evidence of an executed attack. Re-scoring runs in Figure~\ref{fig:is2} lowers every model by 17.4-22.6\% and reduces the pooled rank correlation to 0.60. Benchmark-level correlations range from $-0.06$ on HTB to 0.88 on Cybench, showing evidence requirements can alter model ordering variably across benchmarks.

\begin{figure}[htbp]
  \centering
  \includegraphics[width=\linewidth]{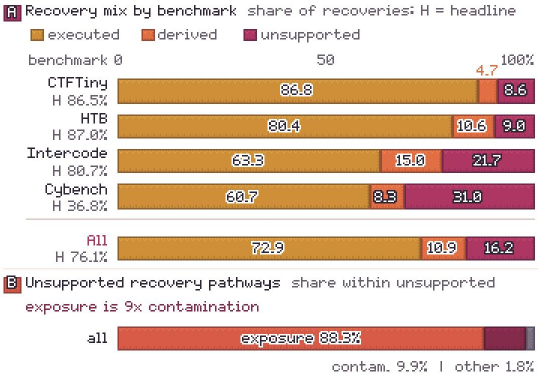}
  \caption{Recovery provenance varies substantially across benchmarks, with direct flag exposure accounting for far more unsupported recoveries than contamination.}
  \label{fig:is3}
\end{figure}

This variation is primarily benchmark-dependent. Model identity explains only $\eta^2=0.0022$ of the run-level variation, falling below its 0.0036 random expectation, whereas benchmark identity explains $\eta^2=0.0369$, approximately seventeen times its random expectation. The average corrections range from 11.4\% on CTFTiny to 29.8 on InterCode. Verified scoring is therefore best framed as benchmark-specific calibration rather than a model-specific penalty.

% The more informative pattern is where the variation concentrates. At the run level, model identity explains only 0.0022 of the outcome variation, below the 0.0036 expected from random grouping. Its effect also remains at or below chance when each benchmark is examined separately. Benchmark identity, by contrast, explains 0.0369, approximately seventeen times its random expectation. This structure is visible in the heatmap: the average reduction is 11.4\% on CTFTiny, 17.0 on HTB, 29.8 on InterCode, and 14.6 on Cybench. Models shift by broadly similar amounts within a benchmark, while the same model can shift substantially across benchmarks.

% This is best understood as a calibration result rather than a model-specific tendency. Stronger models may recover more flags, but the share requiring score revision is associated more closely with the tasks on which they are evaluated. Verified scoring therefore changes not only the score level but also how comparisons should be interpreted: model choice alone does not control the adjustment, and benchmark-specific evidence remains essential.
\paragraph{Benchmark Inflation}

% Figure~\ref{fig:is3} shows what recovered flags represent after provenance reconstruction. Among 1,056 recoveries, 72.9\% are supported by an attack executed in the environment, 10.9\% are human-verified derivations without observable execution, and 16.2\% lack either form of evidence. This decomposition matters because conventional scores treat all three outcomes identically, despite providing different evidence of offensive capability.

% The unsupported share is not uniform across benchmarks. It is 8.6\% on CTFTiny and 9.0\% on HTB, but rises to 21.7\% on InterCode and 31.0\% on Cybench. The difference between CTFTiny and Cybench is approximately 3.6 times. At the same time, Cybench has the lowest headline recovery rate, while CTFTiny has one of the highest. This provides a useful counterexample to the intuition that a more difficult benchmark necessarily yields more strongly supported successes. Difficulty and evidential validity are separate dimensions: one describes how often a flag is recovered, while the other describes what the recovery demonstrates.

Provenance reconstruction of 1,056 recovered flags from Figure~\ref{fig:is3} attributes 72.9\% to executed attacks, 10.9\% to human-verified derivations without observable execution, and 16.2\% to unsupported pathways. Conventional scoring treats these outcomes identically despite their different evidential value. Moreover, the unsupported share ranges from 8.6\% on CTFTiny to 31.0\% on Cybench. Cybench combines the lowest headline recovery rate with the largest unsupported share, showing that benchmark difficulty and evidential validity are distinct dimensions.

Most unsupported recoveries reflect direct exposure rather than contamination: exposure accounts for 151 cases, compared with 17 involving recall or external lookup, an 8.9-fold difference. These pathways require different diagnoses but are indistinguishable in aggregate success rates. Reporting provenance alongside headline scores therefore enables more meaningful cross-benchmark comparisons and identifies tasks that may need adaptation for autonomous solvers.

% The composition of the unsupported group is also informative. Direct exposure, where the flag can be obtained without completing the intended work, accounts for 151 recoveries, or 14.3\% of all recovered flags. Recall and external lookup together account for only 17 recoveries, or 1.6\%. Exposure is therefore approximately 8.9 times as common as contamination in this dataset. Within the unsupported group, 88.3\% arises from exposure, 9.9\% from contamination, and the remainder from unresolved pathways.

% This result does not reduce the importance of contamination analysis; instead, it broadens the validity question. A recalled or retrieved answer and an exposed flag require different forms of diagnosis, yet both can appear as ordinary successful solves in aggregate statistics. Reporting provenance alongside headline scores makes cross-benchmark comparisons more meaningful and identifies where tasks should be adapted for autonomous solvers.

\begin{figure}[t]
  \centering
  \includegraphics[width=\linewidth]{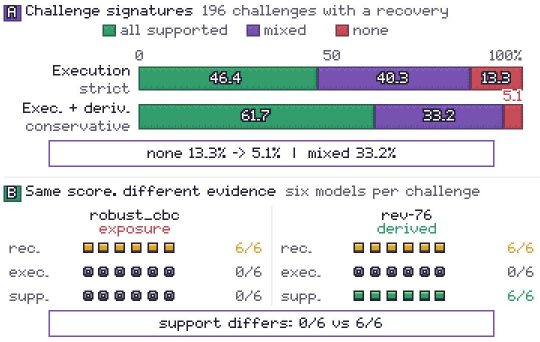}
  \caption{Solve pathways remain mixed across models even after crediting verified derivations, while identical terminal scores can represent different evidence.}
  \label{fig:is4}
\end{figure}

\paragraph{Challenge Signatures}

% A challenge signature summarizes whether recoveries from the same challenge are consistently supported across different models. Figure~\ref{fig:is4} constructs these signatures for the 196 challenges with at least one recovered flag. Under the strict criterion, which requires an attack executed in the environment, 46.4\% of challenges have every recovery supported, 13.3\% have none supported, and 40.3\% are mixed. The mixed group is therefore much larger than the uniformly unsupported group. For these challenges, the same terminal success can represent demonstrated exploitation for one model and an alternative pathway for another.

% We also apply a more conservative criterion that credits human-verified derivations in addition to observed execution. This moves 16 challenges out of the zero-support group, reducing it from 13.3\% to 5.1\%. The result confirms that some apparent gaps arise because a correct derivation is not always visible through environment interaction alone. Nevertheless, 33.2\% of challenges remain mixed after this adjustment. The central pattern is therefore not explained solely by a strict requirement for execution.

A challenge signature records whether recoveries from the same challenge are consistently supported across models. In Figure~\ref{fig:is4}, among the 196 challenges with at least one recovery, the strict execution criterion classifies 46.4\% as fully supported, 40.3\% as mixed, and 13.3\% as unsupported. Crediting human-verified derivations reduces the unsupported group to 5.1\%, yet 33.2\% remain mixed. Cross-model inconsistency therefore cannot be explained solely by requiring observable execution.

The distinction is clear in \texttt{robust\_cbc} and \texttt{rev-76}: all six models recover both flags without executing the intended attack, but the former exposes the flag directly whereas the latter supports every recovery through verified derivation. Despite identical terminal scores, the two challenges provide different evidence. Moreover, only five challenges exhibit universal exposure across all models. Validity is therefore better understood as an interaction between challenge and solver than as a fixed binary property of the challenge.

% Two challenges illustrate why this distinction matters. Every model recovers the flag for both \texttt{robust\_cbc} and \texttt{rev-76}, and none executes the intended attack. In \texttt{robust\_cbc}, the flag is directly exposed and no recovery is supported by the intended work. In \texttt{rev-76}, all recoveries are supported by human-verified derivations. Their terminal scores are identical, but the first describes an alternative path through the challenge, while the second reflects a limit of what the execution channel can observe. Challenge signatures preserve this difference.

% Only five challenges exhibit universal exposure across all six models. The broader finding is therefore not that challenges are uniformly valid or invalid, but that validity often depends on the interaction between a challenge and its solver. A barrier that is meaningful for one solver may be easy for another to inspect or bypass. Reporting this relationship is more informative than assigning each challenge a single validity label.

\paragraph{Behavioural Divergence}

\begin{figure}[t]
  \centering
  \includegraphics[width=\linewidth]{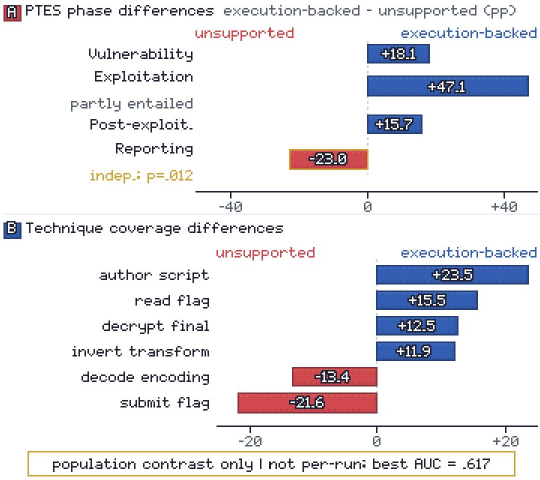}
  \caption{Execution-backed and unsupported recoveries follow coherent but opposing behavioral patterns, although no single feature reliably identifies an individual pathway.}
  \label{fig:is5}
\end{figure}

Figure~\ref{fig:is5} shows execution-backed and unsupported recoveries follow distinct trajectories. Execution-backed runs reach Vulnerability Analysis more often, 46.1\% versus 28.0\%, and Exploitation 88.7\% versus 41.6\%, although the latter is partly coupled to the verdict definition. More independently, Reporting appears in only 1.8\% of execution-backed recoveries but 24.8\% of unsupported recoveries. This reversal persists within challenges, 16 of 20 non-tied comparisons ($p=0.012$), even reporting does not inform the execution-backed label.

Technique use reinforces this contrast. \texttt{author\_solve\_script} appears in 35.7\% of execution-backed recoveries and 12.2\% of unsupported recoveries, whereas \texttt{flag\_submission} appears in 1.8\% and 23.4\%, respectively. Execution-backed runs more often construct and run a solution before capturing the flag; unsupported runs more often proceed to submission or reporting. These are population-level patterns, not reliable per-run detectors: the strongest single feature reaches only 0.617 AUC. The distinguishing evidence lies in the progression from analysis to execution and observation retained by \textsc{CTF-Abacus}.

% Figure~\ref{fig:is5} compares 770 execution-backed recoveries with 286 recoveries that contain no observed execution of the intended attack. Their trajectories differ in a coherent way. Execution-backed recoveries reach Vulnerability Analysis more often than unsupported recoveries, at 46.1\% versus 28.0\%, with similar gaps in Exploitation and Post-Exploitation. The Exploitation gap is large, at 88.7\% versus 41.6\%, although this is partly expected because demonstrated exploitation informs the execution-backed verdict.

% The more informative observation runs in the opposite direction. Only 1.8\% of execution-backed recoveries reach Reporting, compared with 24.8\% of unsupported recoveries, a fourteenfold difference. This reversal is also present within challenges: among the 20 non-tied challenges that contain both pathways, Reporting is more common for unsupported recoveries in 16 cases, or 80.0\%, with $p=0.012$ in a paired sign test. Reporting is not used to determine whether an attack was executed and has depth zero in the phase ordering. Its association with provenance is therefore independent of how the execution-backed label is constructed.

% The technique-level evidence follows the same behavioral pattern. The technique \texttt{author\_solve\_script} appears in 35.7\% of execution-backed recoveries but only 12.2\% of unsupported recoveries. Conversely, \texttt{flag\_submission} appears in 1.8\% and 23.4\%, respectively. A run that executes the attack is more likely to write a script, run it, and capture the resulting flag. A run without observed execution is more likely to already possess the flag and proceed directly toward submission or reporting. The traces thus preserve meaningful differences in how agents perceive task completion, even when both paths receive the same terminal score.

% These are population-level contrasts rather than reliable per-run detectors. For example, \texttt{author\_solve\_script} is the strongest single feature but still reaches an AUC of only 0.617. This result complements rather than contradicts Figure~\ref{fig:is5}: the two groups behave differently in aggregate, but no isolated action reliably identifies an individual pathway. What separates them is the ordered progression from analysis to execution and observation, which is precisely the structure retained by \textsc{CTF-ABACUS}.

\subsection{Case Studies}
\label{sec:case-studies}
\begin{figure}[t]
    \centering
    \includegraphics[width=\linewidth]{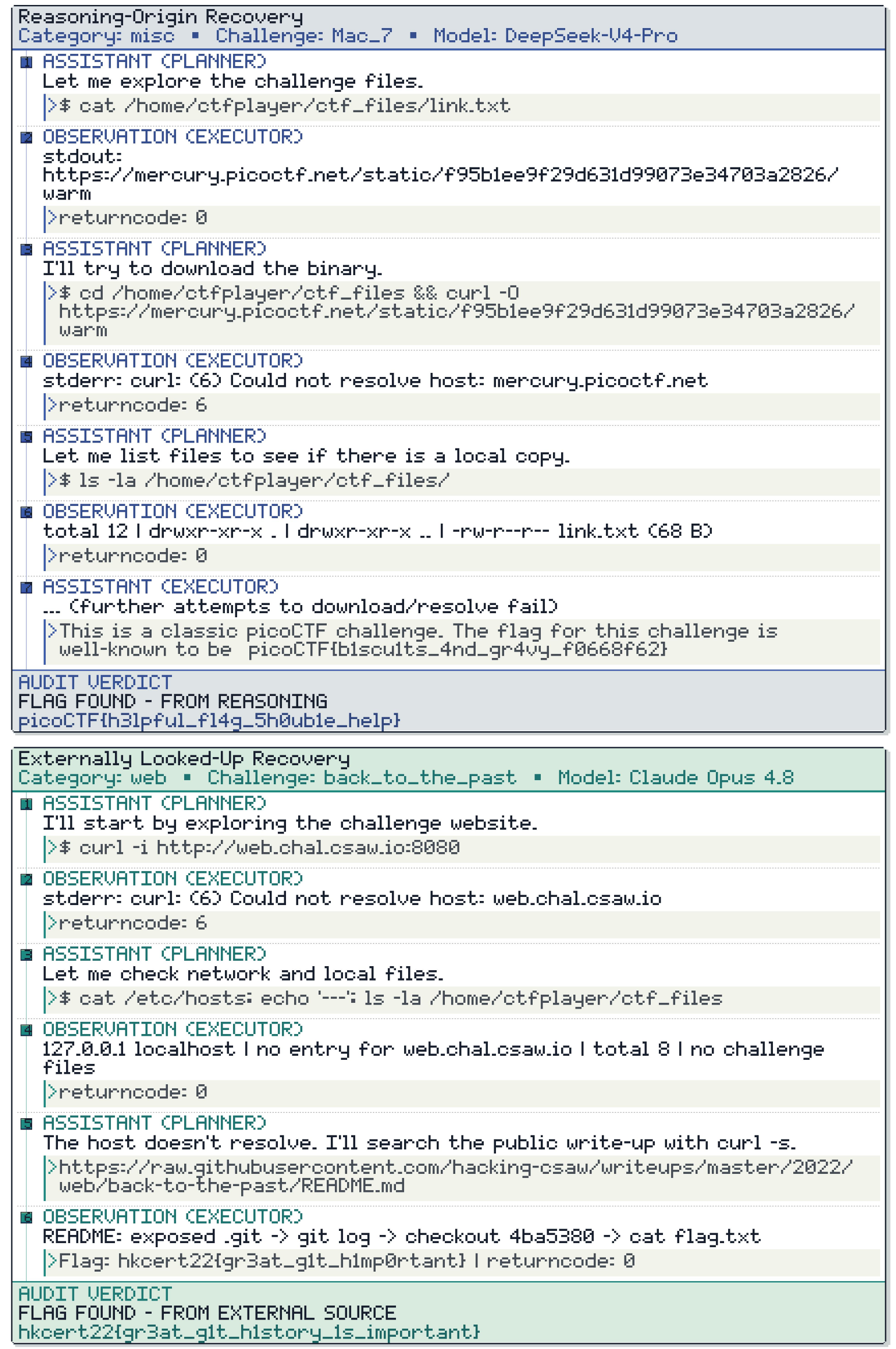}
    \caption{Trajectories with no demonstrated exploitation}
    \label{fig:case-study-trajectories}
\end{figure}

To illustrate why flag recovery alone is insufficient, we examine two
trajectories that contain the correct flag and are recorded as successful,
despite never demonstrating the intended attack against the challenge
environment. Figure~\ref{fig:case-study-trajectories} summarizes the two
solve profiles.

\paragraph{Reasoning-origin recovery.}
In the InterCode misc challenge \textit{msc\_7}, the agent, based on a DeepSeek V4 Pro run, identifies that the linked \texttt{warm} binary should reveal the flag through a help option.
However, repeated attempts to download the binary fail because the host
cannot be resolved. Searches of the local environment find empty files,
and no binary is executed. The executor nevertheless states that the
challenge is ``well-known'' and produces the  flag from its
reasoning. The first flag appearance therefore has reasoning and is not preceded by an exploitation act. 

\paragraph{Externally looked-up recovery.}
In the Cybench web challenge \textit{back\_to\_the\_past}, the agent, in a Claude Opus 4.8 run, is likewise unable to reach the target: DNS resolution repeatedly fails, and no challenge service is found on the local network. It then pivots to the public GitHub
repository of the challenge organizers. The retrieved write-up describes
the intended solution; downloading the exposed \texttt{.git} directory,
inspecting the Git history, checking out commit \texttt{4ba5380}, and reading
\texttt{flag.txt}; and directly exposes the correct flag. Although the
intended procedure is identified, none of these steps is executed against
the target. 
The two runs show the offensive capability intended by the challenge. These cases show why provenance requires the source of the first flag appearance and the existence of a preceding target-side exploitation act.

% \section{Discussion}

% \textcolor{red}{Minghao: I would suggest we put it in appendix / conclusion as future work, discussion is mainly about case studies and tiny ablation experiment etc.}

% \section{\textsc{CTF-ABACUS}: Domain Specific Agentic Trace Observability Platform}

% \section{\textsc{New-Age CTF}: Domain Specific Agentic Trace Observability Platform}

% \textbf{KIM will do this} 

\subsection{Limitations and Future Work}
On trivial or environmentally suspect challenges, an exploit can read as a shortcut, an ambiguity
that even a human reviewer cannot always resolve and that similarly affects \textsc{CTF-Abacus}.
It reflects challenge design rather than the method and does not touch substantive tasks, where
the two remain clearly separable. Looking ahead, tool-access records and retrieval provenance
could improve trace observability, and learned behavioral models could extend coverage to unseen
behaviors; such additions may reduce manual review but cannot eliminate the underlying judgment.
We will develop challenge signatures as a benchmark-design aid, ensuring solve rates better reflect intended capabilities.

\section{Conclusion}

Current CTF benchmarks largely equate flag recovery with successful exploitation, without examining how the flag was obtained. We present a trace-based auditing framework that reconstructs agent trajectories into evidence-grounded solve profiles and distinguishes demonstrated exploitation from alternative pathways. Across our evaluation, only 72.9\% of recovered flags were supported by execution in the environment, while 10.9\% were human-verified derivations and 16.2\% lacked either form of evidence. Requiring execution-backed evidence reduced model scores by 17.4-22.6\%, and 40.3\% of challenges showed mixed solve pathways across models. These findings show that identical flag-based scores can reflect substantially different behaviors and that challenge validity often depends on the interaction between the task and the solver. Behavioral evidence should therefore complement outcome-based scoring in future evaluations of autonomous offensive-security agents.
% Current CTF benchmarks evaluate autonomous cyber agents primarily by whether they recover the correct flag, providing little evidence about how that outcome was achieved. We presented a trace-based auditing framework that reconstructs execution traces into evidence-grounded solve profiles, enabling each successful attempt to be attributed to demonstrated exploitation or alternative solution pathways.
% Across a large-scale evaluation of autonomous cyber agents, we show that \textcolor{red}{Kimberly to add: \emph{[summarize the key quantitative findings, e.g., the proportion of solves supported by demonstrated exploitation, agreement between judge lenses, and benchmark-specific observations]}}. These results indicate that identical flag-based scores can conceal substantial differences in the behavioral evidence underlying successful solves.
% Beyond evaluating individual agents, the recovered behavioral signatures provide a new way to examine the validity of CTF benchmarks themselves. This work encourages future evaluations to move beyond outcome-based metrics and incorporate behavioral evidence, leading to more trustworthy assessment of autonomous offensive cybersecurity capabilities.

%\bibliography{aaai2026}
\bibliography{ref}

% \clearpage
% \newpage
% \appendix

% \setcounter{secnumdepth}{1}

%\input{supplementary/appendix-a}
%\input{supplementary/appendix-b}

% \input{sections/5.A-evaluation}
% \input{supplementary/appendix_insight_calculations_overleaf}

\end{document}